\documentstyle[12pt]{article}
\title{LOW-BACKGROUND EXPERIMENTS \\
WITH HIGH PRESSURE GAS SCINTILLATION \\
PROPORTIONAL DETECTOR \\}

\vspace{2cm}

\author{
~\\
~\\
D.Yu. Akimov\footnotemark[1], A.A Burenkov, \\
V.F. Kuzichev, V.L. Morgunov\footnotemark[2], V.N. Solovov \\
~\\
~\\
~\\
  \small{Institute for Theoretical and Experimental Physics} \\
  \small{RU-117259  Moscow  Russia} \\         }

\date{4-Mar-1997}

\begin{document}
\vbox{%
\maketitle}

\begin{abstract}

A scintillation proportional counter with wavelength shifting fiber readout
filled with Xe or Kr under a pressure of up to 20 atm is proposed for the
low--background experiments on search for dark matter of the Universe and
$2K$--decay of $~^{78}Kr$.

\footnotetext[1]{E-mail : akimov\_d@vxitep.itep.ru}
\footnotetext[2]{E-mail : morgunov@vxitep.itep.ru}
\end{abstract}



\section{Introduction}

\subsection{The dark matter of the Universe}

Astrophysical observation during the last tens years points out on the 
existence of a significant hidden mass in our Galaxy as in other ones
\cite{dm,barnab}. Modern cosmological theories and also the 
baryosynthesis theory
imply the density of the Universe to be near its critical value of
$\rho_{c} = 3 H^2 / 8 \pi G_{N}$,
where $H$ is the Hubble constant, $G_N$ is the gravitational constant.
The value of the density of the matter observable by the astrophysical
methods is about of 1\% of the critical value. The problem of dark matter
consists of two general points at least:

1) The existence of invisible baryonic matter (with the density 
given from the theory of baryosynthesis of $\sim$ 0.03 - 0.08 $\rho_c$)

2) The existence of nonbaryonic dark matter having the density of  
$\rho > 0.2 \rho_c$ which is necessary to account for the observing rotation 
curves of spiral galaxies.

Division the nonbaryonic dark matter on two components: "hot" (30\%; 
massive neutrino) and "cold" (70\%) is implied from the observing
anisotropy of the relic photons and from the modern cosmology (formation
of galaxy clusters). A Majorana particle -- neutralino with a spin of 1/2
predicted in the supersymmetric models is a
possible candidate to the "cold" dark matter \cite{phrev}. 

Modern experiments on searching for dark matter are carried out in all 
directions. Those are the astrophysical observations (gravitation 
microlenses; EROS--1, MACHO, OGLE experiments; see \cite{barnab}), 
the "standard" experiments on measurement of the mass  of
$\nu_e$, $\nu_{\mu}$, and $\nu_{\tau}$ neutrino, and also the search for
"cold" dark matter by means of direct detection of the recoil nuclei in
the interaction of the supersymmetric particle with the detector substance
\cite{dmexp}. Many of the existing low--background detectors are used in
the search for dark matter in parallel to searching for double
$\beta$--decay \cite{beck1, caldw1}.

\subsection{$2K$--decay of $~^{78}Kr$}

Direct observation of the double $K$--capture in $~^{78}Kr$ is actual 
because it would be the first observation of the process of such a type.
Measurements of the decay time with respect to $2K$--capture for a wide
number of nuclei give the information which can be used for more precise
calculation of the nuclear matrix elements for both $2\beta (2\nu)$ and
$2\beta (0\nu)$ decays. Only lower limits of the decay time was obtained
from the experiments on observation of this process and from
geochemistry estimations \cite{bert}. This fact is caused by the
difficulties of the detection of the low--energy characteristic emission.
$~^{78}Kr$ is the most suitable candidate because it is a target and a good 
working medium for a gas proportional counter simultaneously.

\section{Detector}

The detector is based on a cylindrical high pressure gas proportional counter 
with optical readout by means of WLS fibers proposed by us \cite{akim1, akim2}.

With respect to proportional counter with electrical signal readout (charge
collection) a scintillation proportional counter has the following basic
advantages: 

-- An equivalent electronic noise is 5 - 10 times lower owing to the use of 
PMT as a low--noise device. Therefore, the energy threshold can be lowered down 
to $<$ 1 keV region.

-- Theoretical limit of energy resolution is lower. 

-- There is no microphonic noise absolutely.

-- A HV circuit and spectrometric electronics are totally decoupled from 
each other.

The possibility to have a very low energy threshold with the WLS fiber readout 
was shown with the prototype having 25--cm length and 2.2--cm diameter and
filled with Xe at a pressure of up to 8 atm by demonstrating the capability to
record a single electron originated in the sensitive volume \cite{akim2}.
 
The detector (or the module for larger setup) to be a cylinder with a central 
anode wire surrounded with an array of fibers (Fig. 1, 2). 

The fibers with an outer diameter of 1 mm and a diameter of the core of 0.8 mm
are coated with a wavelength shifter by means of vacuum deposition. The fibers
are to be made from a low--radioactive material. 
The fibers and electrode system are fixed at the ends of the cylinder. 
The fibers are divided in
several groups. Each group of fibers is viewed by two photomultipliers from
opposite directions. The signals from the PMT dynodes come to the triggering
unit, and those from the PMT anodes, to spectrometric analysis.

The titanium pipes are welded in the flanges (Ti) to make a connection of each 
group of fibers with the individual photomultiplier (FEU--85) and of the wires 
with the high voltage feedthroughts. Such a design allows one to reduce as 
much as possible the radioactive background from the standard elements 
(photomultipliers, optical windows, and HV feedthoughts) and also gives one 
the possibility to install a passive shield between these elements with their 
constructions and the body of the detector.

In the detector, which is being developed for measurements in low--background 
conditions in the energy range of less than a few tens keV, a near--wall 
1.5--cm gas layer to be shared out from the total inner gas volume by wire 
cathodes. This layer provides active shielding from the $\alpha$ and $\beta$ 
particles and the soft photons of characteristic emission. An electrical 
readout is planned for the anodes located in this layer (with a threshold of a
few tens keV) by means of charge--sensitive preamplifiers coupled with the
anodes. An additional optical readout is possible too. It can be performed with
the following way. If ionization takes place in the shielding layer then only 
one or two groups of fibers will be irradiated maximally. Otherwise, if the
light signal comes from the central wire, i. e. the ionization takes place in
the volume inside the cathodes (fiducial volume), the fibers of all groups will
be irradiated more uniformly. Simultaneous appearing of light signals in the
fiducial volume and shielding layer will give the sharply nonuniform activation
of the fiber groups too. Thus, measuring the light distribution among fiber
groups one can selects those events which take place only in the fiducial
volume. The events taking place in the near--end region of the counter can be
rejected by the relation of the value of signals coming from the opposite 
ends of fibers (a
light attenuation length in the fiber is $>$ 1 m). In addition, it can be
made also with the use of fibers forming a ring--type wisp and located in
the end wall of the counter in the close radial vicinity to the central anode.
The signal in these fibers can appear only for the near--end events. 

Two options of the detector (module) is considered.

In the first one the body of the detector is manufactured from a 
low--background material (3-mm Ti with a contamination of U/Th $< 10^{-9}$). 
The gas pressure can be of up to 20 atm. 

In another version the body to be made from scintillator (Fig. 3). It is known
that a plastic scintillator can be purified to $10^{-9}$ g/g of  U/Th
contamination \cite{barab1}. The scintillator on the basis of acrylic to be
chosen taking into account the demands of mechanical strength. Moreover, the
acrylic is an excellent radio--pure material, its purity can be of down to  
$10^{-12}$ g/g of U/Th \cite{brodgini}. 
The inner surface of the cylinder to be coated with a specular aluminium layer
(which also plays a role of the cathode). 
The outer surface of the cylinder to be covered with a WLS fibers
for collection of the scintillation light. Thus, the body of the counter is
made to be active, and the events of $\alpha$ or $\beta$--decay followed by 
the emission of $\gamma$--ray, which cause a background in the detector 
sensitive volume, are effectively rejected. 
A part of the Compton--scattered $\gamma$--rays in Xe or Kr is detected too. 
Such a method of light collection allows one to record efficiently the 
scintillation in the detector wall  corresponding to the energy deposition of 
$>$ 0.5 MeV. 
However, for such a design the maximum
pressure to be reduced to 10 atm and the thickness of the wall to be increased
to $\sim$ 1 cm (the mass of the plastic body equals approximately to the mass
of the Ti one). As a consequence, the mass of the sensitive substance is less
by a factor of 2. Nevertheless, additional anticoinsidences allows one to
reduce significantly the detector background. 

The radiopurity of all materials is planned to be measured with the ITEP 
low--background Ge--detector setup \cite{starost}. 

\section{Three experimental stages}

\subsection{Stage I. Building ~~~ the ~~~ laboratory ~~~ setup for 
experiment on search for $2K$--decay of $~^{78}Kr$ and further R\&D in 
low--background conditions}

Basing on the R\&D carried out with the scintillation proportional 
counter and described in \cite{akim1, akim2} we are ready to start now 
building the setup for 
the specific experiment on $2K$--capture in $~^{78}Kr$ \cite{barab2}. This
setup, also, to be a laboratory prototype of the setups of stages II and III
(see below). This stage is necessary for study and measurement of the real
background conditions and working out the technique of shielding and the
triggering system for a large--scale detector. Using this setup we are
planning to outwork a technique of recording the signals from the events with
energy $<$ 1 keV and methods of background reduction down to a few
events/keV/kg/day.

As pointed out in \cite{barab2}, a signature of the event of
$2K$--capture in  $~^{78}Kr$ is simultaneous detection of two $\gamma$--rays of 
characteristic emission of $~^{78}Se$ atom having equal energies ($\sim$ 12
keV). The detector to be a cylinder with an inner diameter of $\sim$ 120 mm and
a length of $\sim$ 450 mm and to be installed in a scintillator active shield
(anti--Compton) and a passive shield made of lead and borated polyethylene. With
a near--wall layer and a near--flange one fixed at $\sim$ 1 cm and $\sim$ 5 cm,
respectively, the value of fiducial volume of the detector to be 3.3 liter.
The total volume of the detector with inlet--outlet pipes to be about 7 liter. 
The 100--g of
$~^{78}Kr$ isotope with an enrichment of 94\% is planned to be used in
experiment. With this amount the pressure to be about 5 atm, and the mass of
the gas in the fiducial volume to be $\sim$ 50 g. The sensitivity of the
experiment expected with this amount of isotope and the background level of
$\sim$ 2 events/keV/kg/day (the value of the background achieved in the ITEP
ground low--background laboratory with the use of such a shield and the
Ge--detector \cite{starost2}) to be $\sim 10^{22}$ years for $T_{1/2}$ of 
$~^{78}Kr$ 2K-capture.  

\subsection{Stage II. Setup for DM search}

At the second stage the manufacture and operation in a real background 
condition of the detector with a mass of $\sim$ 10 kg of Xe and more is 
planned. 

The setup to be consisted of the separated cylindrical modules (Fig. 4). The 
cylinder has an inner diameter of 15 cm, a wall width of 3 mm (Ti), and a
length of 1.2 m. The detector to be filled with Xe gas. The mass of Xe in the
total detector volume and in the fiducial volume is 2.7 kg and 1.35 kg,
respectively. 

Xe is chosen as a working medium owing to the following reasons:

-- It has the high cross--section of spin--independent interaction (because of 
the large number of nucleons in the Xe nucleus) and also it has the isotopes
with a spin 1/2 ($~^{129}Xe$; 26.4\%) and 3/2 ($~^{131}Xe$; 21.2\%) in the
natural Xe, which take part in spin--dependent interaction with dark matter
particles.

-- There are no long--life radioactive isotopes.

-- High degree of purification of large amounts from radioactive contaminations 
is possible.

-- It has a high cross--section of absorption of the low--energy photons 
(E $<$ 20 keV). This fact allows one to reject effectively the characteristic 
emission from the detector body by means of the near--wall active layer of gas.

The event is considered to be a useful one if it appears in the fiducial 
volume of the one module only, looks like point--like energy deposition in the 
range 0.2 - 100 keV, and is not accompanied with the signal (or signals) 
from the shielding layer of this module or any signals from the other modules 
in the time interval $\pm$ $T_{dr}$ ($T_{dr}$ = 60 $\mu$s is the total 
charge collection  time in the counter).

The low--background experiment on dark matter search to be carried out in an 
underground laboratory. The one of the possible places where this experiment
is considered to be carried out is the massive shield of the existing ITEP
setup in Gran Sasso for $2\beta$--decay search (Pb + high--purity Cu; Fig. 4). 

The detector will be sensitive to the SUSY particles with a mass of 
30 - 200 GeV and an upper limit of rate of $\sim$ 1 event/kg/day
for the threshold on energy of recoil nuclei of 5 keV.

The value of 5 keV is a rough estimation 
of a real threshold because the quenching factor is not known exactly for 
20-atm Xe and it is  planned to be measured by means of calibration with 
neutrons.  Thus, the detector is competitive with low--temperature 
bolometers being developed now for dark matter search \cite{absm} because it 
has low energy threshold and significantly exceeds they by mass. It also 
competitive to the liquid Xe detectors proposed in \cite{cline1}

\subsection{Stage III. Large--scale detector}

The stage of a large--scale detector is considered here as a conceptual design
only  because technical difficulties and the price of such a setup will be 
significantly higher than those for the second stage. On the other hand, the 
superior characteristics with respect to those of the existing or planning 
detectors for the same physical goals can be achieved. 

The detector to be a large cylinder with a diameter of 2 m and a length of 
2 m filled with a working gas under a pressure of 20 atm. Hexagonal wire 
cells with a size of $\sim$ 10 cm are installed inside the cylinder. Each cell 
has a central wire anode. 

The body of the detector to be manufactured from the acrylic activated with 
scintillator and has an outward bandage made of superstrength carbon fibers.
A modern spacecraft producing plant is capable to manufacture this construction.
The layer of WLS fibers for readout of the scintillation signals originated  
in the wall is located inside the detector and separated with a light--tight 
film from the layer of fibers for readout of the electroluminescent signals. 

The detector to be placed in the tank filled with a liquid scintillator. The 
whole volume of the scintillator with a thickness of more than half of meter 
is viewed by PMTs and is used as an active shielding from the outer radioactive 
background and for anticoinsidences for suppression the background originated 
from the Compton interactions in the sensitive volume. 

The high degree of purity, which can be achieved for the liquid scintillator, 
and its sensitivity over the whole volume allows one to have the low 
background from the decays in the scintillator itself. The high thickness of
the active shield provides an effective attenuation of the 
$\gamma$--rays coming from the outside of the setup. The last allows one to 
reduce the outer background by a factor of about $10^3$. 

The working substance of such a detector to be deep--purified from the 
radioactive contaminations. For instance, for Xe the purification from 
$~^{85}Kr$ isotope down to a level of better that $10^{-8}$ g/g is required.
The total mass of Xe in such a detector is about 700 kg. 

The detector can be utilized for the following physical experiments: 

-- search for dark matter with the sensitivity of 0.1 event/day/kg;
   with a mass of detector mentioned above it is possible to observe an annual 
   variations of this rate.

-- measurement of neutrino magnetic moment with the use of tritium low--energy 
antineutrino source (100 MCi, \cite{korn}) down to the value of 
$3 \times 10^{-12} {\mu}B$ \cite{morgunov}. 

-- measurement of the decay time of $~^{124}Xe$ isotope with respect to double 
$\beta$--decay up to $T_{1/2}$ $\sim 3 \times 10^{24}$ years;

-- measurement of neutrino weak interaction, including a coherent scattering 
of the reactor antineutrino on nucleus.

\section{Background}

Computer simulation of the background in a low--background detector is a very 
difficult and sometimes insoluble task. This fact is a consequence of 
unpredictable phenomena such as, for example, the presence of a very low 
amount of the one of Eu isotopes in the working surface of the NEMO-II 
detector \cite{barpriv}. 

There are the following background sources in the energy range of 0.2 - 100 keV 
for the detector proposed: 

1. The cosmic ray muons penetrated to the experimental hall.

The contribution of this background is insufficient because the muons give the 
energy deposition higher than the energy range considered. Therefore, they 
contribute only to the total counting rate of the detector.

2. The neutrons originated in the surrounding stocks and also in the 
construction of the detector as a result of the radioactive decays caused
by cosmic rays.

We don't present here the estimation of the background of this type due to the 
complicity of simulation. It is supposed to be not high.

3. The $\gamma$--rays from the radioactive decays both in the bodies of the 
setup and in the nearest surrounding of it. 

According to our Monte Carlo simulations the spectrum of the background is 
flat with the level of $\sim$ 1 event/keV/kg/day in the energy range 
considered (for the $10^{-9} g/g $ of $U/Th$). 

4. The electrons from $\beta$--decays of the elements of U/Th decay series 
emitted from the inner part of the detector body. 

The near--wall layer with a thickness of 1.5 cm separated from the sensitive 
volume with shielding cathodes allows one to record the signals from these 
electrons, and consequently, to reduce the background of this type.

5. The bremsstrahlung radiation from the electrons mentioned in the item 4.

The contribution of this background is taken into account in the item 3.

6. The electrons and gamma--rays originated from the radioactive 85 Kr in 
the working medium (Xe). 

The background of this type is the most substantial one. A deep purification 
of the working medium is required for its reduction. For the Kr (with a 
natural isotopic composition) concentration in Xe of $\sim 10^{-8}$ the 
contribution of the lower part of the electron spectrum to the background 
will be $\sim$ 1 event/keV/kg/day.

\section{Conclusion}

A high pressure gas scintillation proportional detector with a readout by 
means of wavelength shifting optical fibers can be used for building massive 
detector with a low energy threshold. Monte Carlo 
simulation shows that the level of background from the detector body with a 
radiopurity of $10^{-9}$ g/g U/Th to be $\sim$ 1 event/keV/kg/day for 20--atm 
pressure. This level can be reduced by means of the use of an "active" 
detector body made of scintillator. 

The proposed detector could be used for measurement of 
$~^{78}Kr$ $2K$--decay time of up to $10^{-22}$ years, 
dark matter search with a low energy threshold and high sensitivity, 
and also the measurement of neutrino magnetic moment with an artificial 
source and coherent $\nu N$--scattering.

\section{Acknowledgements}

The authors are grateful M.V. Danilov for the great interest to this work and
the A.S. Barabash for the fruitful discussions.


Fig. 1. Layout of electrodes and fibers inside the cylindrical body of the 
counter. 1 -- counter body, 2 -- WSL fibers, 3 -- wire anodes, 4 -- wire 
cathodes, 5 -- central anode.\\
\\

Fig. 2. General view of the detector (module) proposed for 1.35 kg Xe.\\
\\

Fig. 3. Layout of electrodes and fibers inside the cylindrical body of the 
counter made of scintillator. 1 -- outer WLS fibers,
2 -- counter body, 3 -- WSL fibers for readout of electroluminescence, 
4 -- wire anodes, 5 -- wire cathodes, 6 -- central anode.\\
\\

Fig. 4. Layout of modules in the shield. 1 -- Pb--shield, 2 -- Cu--shield,
3 -- detector modules.\\
 
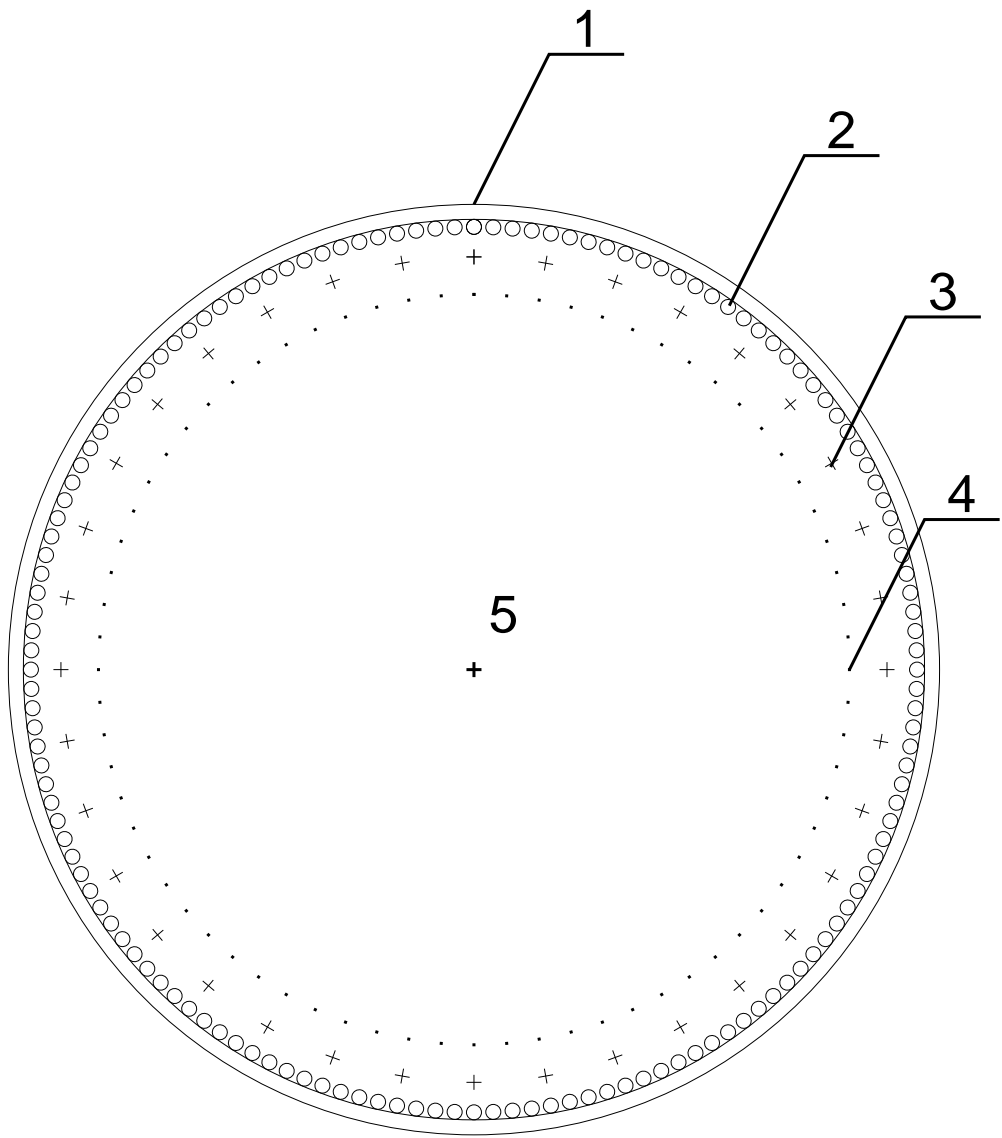
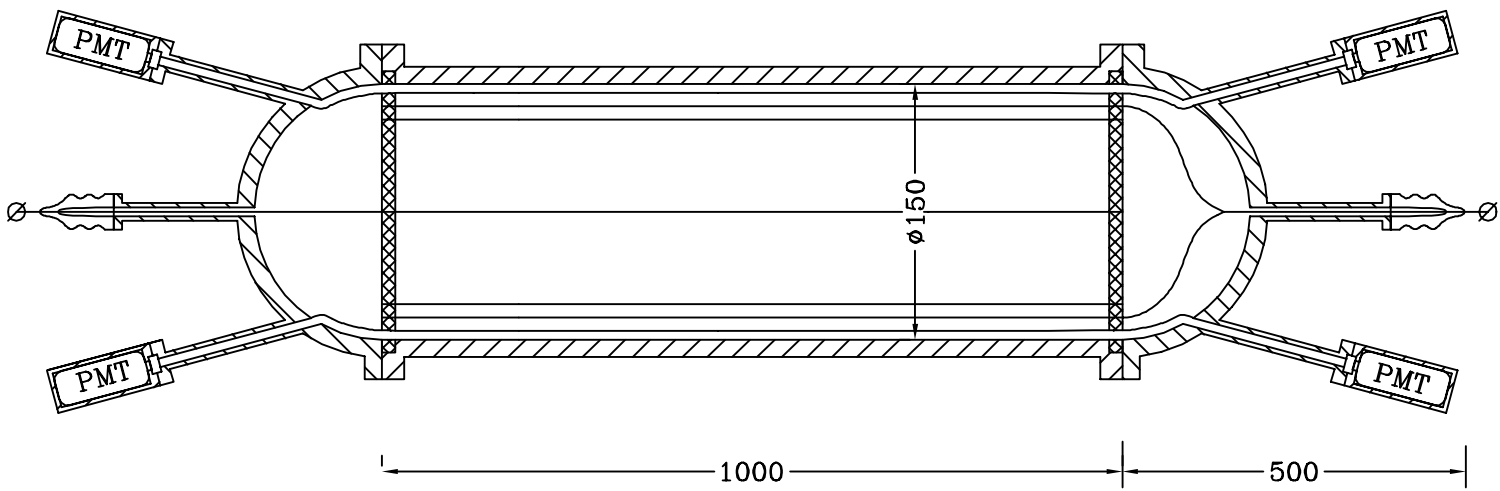
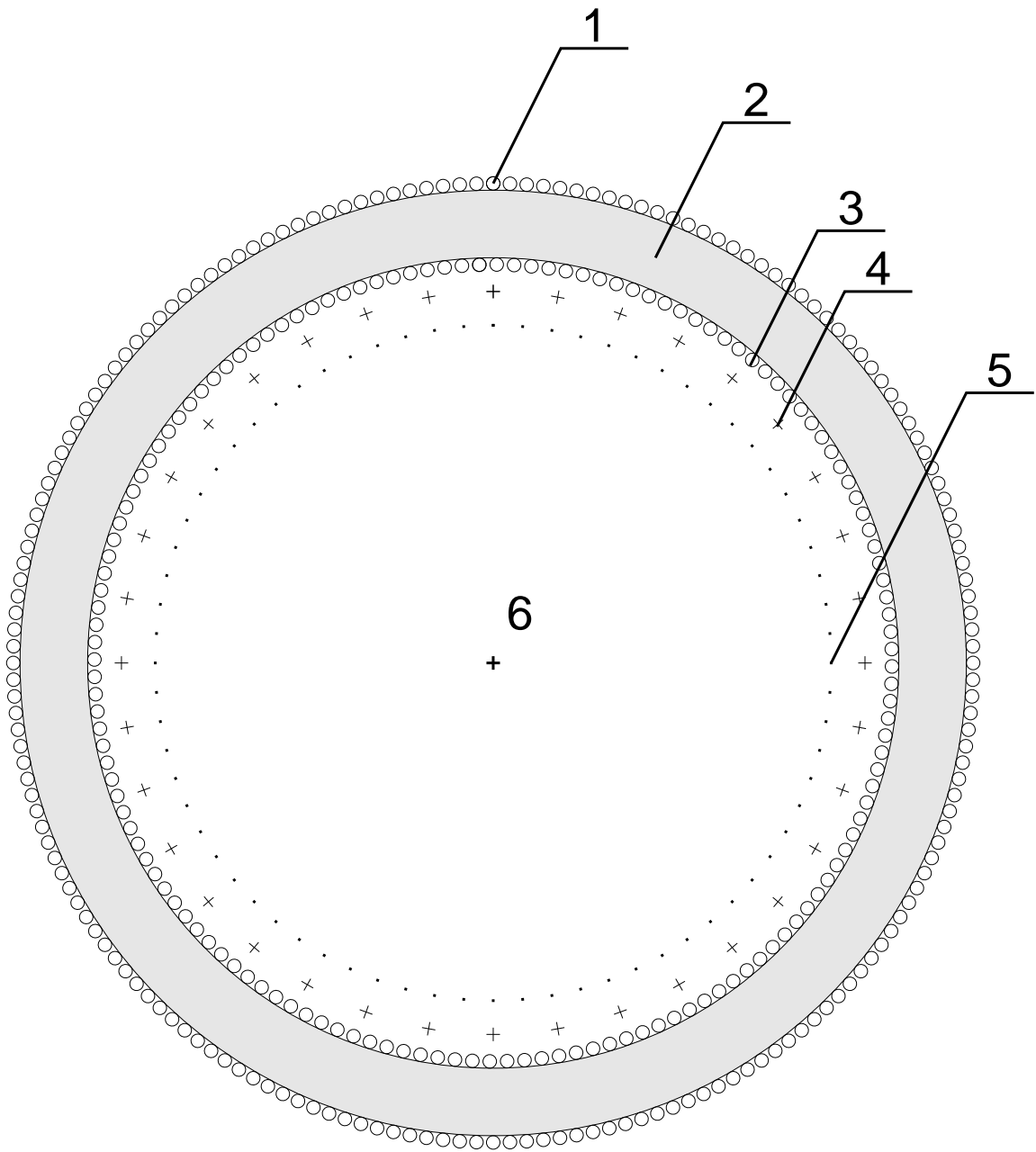
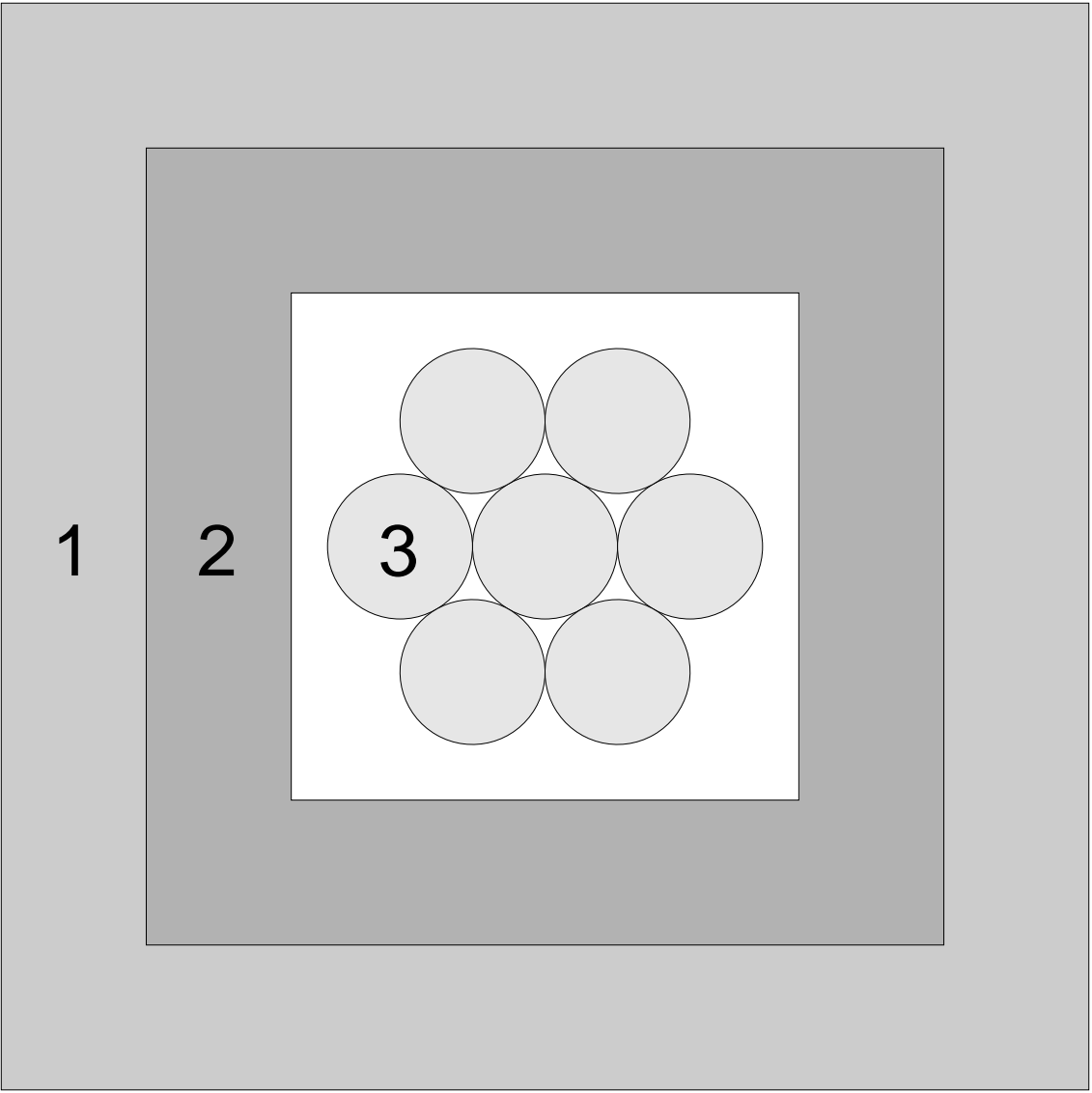
\end{document}